\documentclass[aps,prb,twocolumn,superscriptaddress,amsmath,amssymb,amsfonts,showpacs]{revtex4}

\usepackage{graphicx}
\usepackage{bm}

\newcommand{\bea}{\begin{eqnarray}}

\newcommand{\eea}{\end{eqnarray}}

\newcommand{\beq}{\begin{equation}}

\newcommand{\eeq}{\end{equation}}

\newcommand{\lav}{\langle}

\newcommand{\rav}{\rangle}

\newlength{\textwidthm}

\begin{document}

\def \tr{{\mbox{tr~}}}
\def \ra{{\rightarrow}}
\def \ua{{\uparrow}}
\def \da{{\downarrow}}
\def \ba{\begin{array}}
\def \ea{\end{array}}
\def \nn{\nonumber}
\def \l{\left}
\def \rg{\right}
\def \half{{\frac{1}{2}}}
\def \etal{{\it {et al}}}
\def \cH{{\cal{H}}}
\def \cE{{\cal{E}}}
\def \cK{{\cal{K}}}
\def \cM{{\cal{M}}}
\def \cN{{\cal{N}}}
\def \cQ{{\cal Q}}
\def \cI{{\cal I}}
\def \cV{{\cal V}}
\def \cD{{\cal D}}
\def \cP{{\cal P}}
\def \cF{{\cal F}}
\def \cZ{{\cal Z}}
\def \cC{{\cal C}}
\def \cO{{\cal O}}
\def \cU{{\cal U}}
\def \bS{{\bf S}}
\def \bI{{\bf I}}
\def \bL{{\bf L}}
\def \bG{{\bf G}}
\def \bQ{{\bf Q}}
\def \bK{{\bf K}}
\def \bR{{\bf R}}
\def \bD{{\bf D}}
\def \be{{\bf e}}
\def \br{{\bf r}}
\def \bu{{\bf u}}
\def \bp{{\bf p}}
\def \bq{{\bf q}}
\def \bk{{\bf k}}
\def \bz{{\bf z}}
\def \bx{{\bf x}}
\def \bv{{\bf v}}
\def \bn{{\bf n}}
\def \bw{{\bf w}}
\def \bpsi{{\overline{\psi}}}
\def \tJ{{\tilde{J}}}
\def \K{{\kappa}}
\def \W{{\Omega}}
\def \lam{{\lambda}}
\def \L{{\Lambda}}
\def \a{{\alpha}}
\def \t{{\theta}}
\def \T{{\Theta}}
\def \b{{\beta}}
\def \g{{\gamma}}
\def \D{{\Delta}}
\def \d{{\delta}}
\def \w{{\omega}}
\def \s{{\sigma}}
\def \f{{\phi}}
\def \F{{\Phi}}
\def \x{{\chi}}
\def \e{{\epsilon}}
\def \h{{\eta}}
\def \G{{\Gamma}}
\def \z{{\zeta}}
\def \nd{{^{\vphantom{\dagger}}}}
\def \yd{^\dagger}
\def \iuno{{\bf \hat{i}_1}}
\def \idos{{\bf \hat{i}_2}}
\def \itres{{\bf \hat{i}_3}}
\def \juno{{\bf \hat{j}_1}}
\def \jdos{{\bf \hat{j}_2}}
\def \jtres{{\bf \hat{j}_3}}
\def \hr{{\bf \hat{r}}}
\def \hn{{\bf \hat{n}}}
\def \hj{{\bf \hat{j}}}
\def \hi{{\bf \hat{i}}}
\def \he{{\bf \hat{e}}}
\def \hz{{\bf \hat{z}}}

\title{Lenosky's energy and the phonon dispersion of graphene}

\author{S.~Viola~Kusminskiy, D. K. Campbell, A. H. Castro Neto}
\affiliation{Department of Physics, Boston University, 590 Commonwealth Ave., Boston, MA 02215}

 
\begin{abstract}
We calculate the phonon spectrum for a graphene sheet resulting from the model proposed by T. Lenosky {\it et al.} (Nature {\bf 355}, 333 (1992)) for the free energy of the lattice. This model takes into account not only the usual bond bending and stretching terms, but captures the possible misalignment of the $p_z$ orbitals. We compare our results with previous models used in the literature and with available experimental data. We show that while this model provides an excellent description of the flexural modes in graphene, an extra term in the energy is needed for it to be able to reproduce the full phonon dispersion correctly beyond the $\Gamma$ point.
\end{abstract}


\maketitle

\section{Introduction}
The phonon dispersion of graphite is a reccurrent topic in the literature, and the last years have seen a revival of interest due to the successful isolation of a single layer of graphite, graphene \cite{Geim_review}. Due to the very weak interaction between the planes, the phonon spectra for graphene and graphite are essentially the same, except at very low frequencies where the splitting of the out of plane acoustic mode in graphite is noticeable \cite{Rubio04}. In general the theoretical models most successful in describing phonon dispersions are first principle ones, and graphite is not the exception. However simple, analytical models that would give a good qualitative description of the system are highly desirable. These models are valence force models and usually require many free parameters to give accurate results. The generally accepted as best fitting valence force model for graphite is the one given in Ref. \onlinecite{Jishi82}, in which one considers interactions up to four nearest neighbors and 20 fitting parameters. More recently a five nearest-neighbor model was shown to give a very good fit to new available experimental data \cite{Maultzsch07}. Also a valence force model based in the symmetries of the lattice was proposed \cite{Falkovsky08}. Simpler valence force models in which only nearest neighbor interactions are considered with only two free parameters (these models treat only the in plane modes) are given by Kirkwood \cite{Kirkwood39}, in which the elastic energy cost of stretching and bond bending are considered, and by Keating \cite{Keating66}, in a model that takes into account the symmetries of the lattice. These models can be extended to include the out of plane modes by adding a ``dangling bond'' term \cite{Kitano56,Lobo97}. The Kirkwood model is already a harmonic model, while the Keating model includes up to quartic order terms. The expansion up to second order of the Keating model gives an effective model very similar to the Kirkwood model, although it outperforms it slightly. For this reason in this paper we will take as a reference the extended Keating model --- with the dangling bond term --- as presented in Ref. \onlinecite{Lobo97}.

In Ref. \onlinecite{Haas92} it was shown that for improving the accuracy of either the Kirkwood or Keating model, it is necessary to take into account the overlap of the $\pi$ and $\s$ orbitals due to bending. In Ref. \onlinecite{Haas92} this was done by considering a full quantum mechanical model by use of a tight-binding formalism. On the other hand, in the work by Lenosky {\it et al.} \cite{Lenosky92}, the authors propose a microscopic form for the energy of a graphene sheet that takes into account this overlap by considering the energy cost of having misaligned normal vectors to the graphene membrane. The original model was applied to treat the energetics of negatively curved graphene structures, denominated schwarzites and it was later extended to describe the elastic properties of nanotubes \cite{Zhou00}. In this work we study this proposed free energy and calculate the resultant phonon dispersion, comparing with the available experimental data. We will show that the terms involving the misalignment of the normals flatten the dispersion of the optical modes at the $\Gamma$ point, a flattening that is observed experimentally --- leaving aside, of course, the Kohn anomaly of the longitudinal optic (LO) model \cite{Pisana07}. In particular the out-of-plane modes ZA (flexural acoustic) and ZO (flexural optic) are very well fitted. Graphene is the experimentally realized example of a polymerized membrane and these out-of-plane modes play a fundamental role since it is the non-linear coupling between them and the in-plane modes which stabilizes the flat phase \cite{NelsonBook}, and they are directly related to the rippling observed in the graphene sheets \cite{Meyer07}.  We will see however that, even though Lenosky's energy describes qualitatively the phonon dispersion near the $\Gamma$ point, it gives the wrong ordering in energy for the modes at the $K$ point. We circumvent this by adding an extra term to Lenosky's original formulation, the bond bending term, which is present in both Kirkwood and Keating models but not in Lenosky's.

\section{The Model}
\begin{figure}
 \centering
 \includegraphics[width=4cm,bb=17 40 581 709]{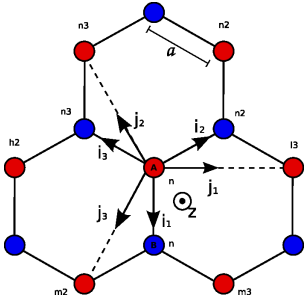}
\caption{Graphene lattice with the notation used throughout the text.}
 \label{fig:labels}
\end{figure}

In graphene the carbon atoms form a honeycomb lattice as shown in Fig. \ref{fig:labels}. This configuration is due to the symmetry of the bonds between the carbon atoms: their $s$ and $p$ orbitals hybridize in the form $sp^2$ which results in $\s$ bonds contained in the graphene plane, while the remaining $p_z$ orbital is perpendicular to the plane and forms the $\pi$ and $\pi^*$ bands \cite{AntonioRMP}. The proposed form for the energy given by Lenosky and collaborators in Ref. \onlinecite{Lenosky92} is given by the following expression:
\beq\label{UL}
\begin{split}
 \cU_L&=\frac{\e_0}{2}\sum_{<ij>}\l(|\br_{ij}|-|\be_{ij}|\rg)^2+\e_1\sum_i\l(\sum_{<j>}\hr_{ij}\rg)^2\\
&+\e_2\sum_{<ij>}\l(1-\hn_i\cdot\hn_j\rg)+\e_3\sum_{<ij>}\l(\hn_i\cdot\hr_{ij}\rg)\l(\hn_j\cdot\hr_{ji}\rg)
\end{split}
\eeq
where $\br_{ij}$ is the vector that points from atom $i$ to atom $j$ in the lattice, $\hr_{ij}=\frac{\br_{ij}}{|\br_{ij}|}$ and $\hn_i$ is the normal vector to the plane determined by the three nearest neighbors of atom $i$ (we will call this loosely the ``normal at atom $i$''). $\br_{ij}=\be_{ij}$ for the undeformed lattice, where $\be_{ij}$ is a unit vector and we have absorbed the lattice constant $a=1.42 \text{\r{A}}$ in the definition of the elastic constants $\e_n$. The indices $i$, $j$ run over the $N$ atoms of the lattice and $<ij>$ denotes nearest neighbors. According to Ref.\onlinecite{Lenosky92}, the first two terms in expression \eqref{UL} correspond to bond stretching and angle bending respectively. We will however refer to this second term as the ``dangling bond'' term since it is essentially the same term that was introduced as such in Ref. \onlinecite{Lobo97} ~\cite{note0}. The last two terms take into account the energy cost of orbital overlapping due to rippling and are ``non-local'' in  the sense that they involve more than nearest-neighbor terms as we will see below. The first of these two terms can be though as related to the $\pi$-$\pi$ orbital overlap, since it involves the scalar product between normal vectors at neighboring atoms, while the second is the projection of the normal onto the bonds and therefore related to the $\s$-$\pi$ orbital overlap.

\section{Dynamical Matrix}
For obtaining the phonon modes we derive the dynamical matrix determined by \eqref{UL} {\it via} a harmonic expansion. If we allow for small displacements of the atoms with respect to their equilibrium positions in the lattice, we can write $\br_{ij}=\be_{ij}+\bu_j^{Y}-\bu_i^{X}$, where $\bu_i^X$ is a small displacement of the $i^{th}$ atom in sublattice $X$, with $X=A$, $B$, and similarly for $\bu_j^Y$ ~\cite{note1}. Up to second order in the displacements we get that expression \eqref{UL} can be written as $\cU_L=\cU_L^{\e_0}+\cU_L^{\e_1}+\cU_L^{\e_2}+\cU_L^{\e_3}$, as we detail below. The stretching energy is given by the usual expression
\beq
\cU_L^{\e_0}=\sum_{X\neq Y}\sum_{\lav ij\rav}\frac{\e_0}{2}\l[\be_{ij}\cdot\l(\bu_j^Y-\bu_i^X\rg)\rg]^2\,.
\eeq
The dangling bond term can be written in compact form by use of the dyadic notation as
\begin{widetext}
\beq
\cU_L^{\e_1}=\e_1\sum_{X\neq Y}\sum_n\l[\sum_{\a(n)}\l(\bu_{\a(n)}^X-\bu_n^Y\rg)\l(\hj_\a^2+\hz^2\rg)\l(\bu_{\a(n)}^X-\bu_n^Y\rg)+\sum_{\a(n)\neq\b(n)}\l(\bu_{\a(n)}^X-\bu_n^Y\rg)\l(Id-\frac{\hi_\a\hi_\b}{2}-2\hi_\a^2\rg)\l(\bu_{\b(n)}^X-\bu_n^Y\rg)\rg]\,.
\eeq
\end{widetext}
We will denote from now on nearest neighbors with Greek indices. With this notation, $\bu_{\a(n)}^X$ indicates the displacement of the $\a^{th}$ neighbor ($\a=1$, 2, 3) of atom $n$. The supra-index indicates that the displacement corresponds to an atom in sublattice $X$. The unit vectors $\hi_\a$ and $\hj_\a$ are as indicated in Fig. \ref{fig:labels} ~\cite{note2} and $Id$ is the 3$\times$3 identity matrix.  

The expansion for the terms that involve the normals to the graphene plane is more involved. Assuming a labeling as in Fig. \ref{fig:labels} we can write $$\bn_n^A=\l(\br_{nn2}-\br_{nn}\rg)\times\l(\br_{nn3}-\br_{nn}\rg),$$ which is a vector of length equal to the area of a unit cell and it is normal to the $n^{th}$ atom of sublattice $A$, and analogous expressions hold for $\bn_{\a(n)}^B$.  Using the explicit expressions for $\br_{ij}$ in terms of the atoms' small displacements it can be shown that $$\bn_i^{A(B)}=\frac{3}{2}\sqrt{3}\bk+\sum_{}\l(\bu_{1(i)}^{B(A)}\times\bu_{2(i)}^{B(A)}{\substack{-\\(+)}}\sqrt{3}\juno\times\bu_{1(i)}^{B(A)}\rg),$$ where the sum is over cyclic permutations of the nearest neighbor index. Therefore the unit vector normal to atom $n$ in sublattice $X$ is given by $\hn_i^X=\frac{\bn_i^X}{|\bn_i^X|}$. It is straightforward to retain the quadratic terms of expressions of the type $\hn_i^X\cdot\,\hn_j^Y$, however for $\l(\hn_i^X\cdot\, \hn_i^X\rg)^{-1/2}$ we have to use a multivariable Taylor expansion. The expansion depends on nine variables since $|\bn_i^X|^{-1}=\l(\bn_i^X\cdot \,\bn_i^X\rg)^{-1/2}$ depends on the three 3D vectors $\bu_{\a(i)}^Y$. If we define the function $f^X(\bu_{\a(i)}^Y)=|\bn_i^X|^{-1}$ we can write
\beq\label{taylor}\nonumber
f^X\l(\bu_{\a(i)}^Y\rg)=\l.\sum_j \frac{1}{j!}\l[\sum_{\x=1}^3\sum_{m=x}^z u_{\x m}^Y \partial_{\x m}'\rg]^j f(u_{\x m}^{'Y})\rg|_{u_{\x m}^{'Y}=0}
\eeq
where $u_{\x m}^Y$ is the m$^{th}$ component of the small displacement of an atom in sublattice $Y$ which is the neighbor $\x$ of atom $i$ in sublattice $X$. From the definition of the normal vectors it is evident, as we mentioned previously, that terms that involve the product of normals at neighboring sites contain nevertheless displacements that go beyond nearest-neighbor interactions.  Putting all together we obtain for the $\pi$-$\pi$ and $\pi$-$\s$ overlap energy cost the following expressions, respectively:
\begin{widetext}
\bea
\cU_L^{\e_2}&=&\e_2\frac{2}{9}\sum_{X\neq Y}\sum_{\lav ij\rav}\l\{\sum_\a\l[\hz\cdot\l(\bu_{\a(j)}^X+\bu_{\a(i)}^Y\rg)\rg]^2-\sum_{\a,\;\nu>\a}\l[\hz\cdot\l(\bu_{\a(j)}^X+\bu_{\a(i)}^Y\rg)\rg]\l[\hz\cdot\l(\bu_{\nu(j)}^X+\bu_{\nu(i)}^Y\rg)\rg]\rg\} \, ,\\
 \cU_L^{\e_3}&=& \e_3\sum_{X\neq Y}\sum_{n}\l\{\hz\cdot\l[\bu_n^X-\frac{1}{3}\sum_\lam\bu_{\lam(n)}^Y\rg]\hz\cdot\sum_\a\l[\bu_{\a(n)}^Y-\frac{1}{3}\sum_\lam\bu_{\lam(\a)}^X\rg]\rg\}\, .
\eea
\end{widetext}
From these expressions we can see that $\cU_L^{\e_2}$ is related to acoustic modes while $\cU_L^{\e_3}$ involves relative displacements of the $A$ and $B$ sublattices and then it is related to the optical modes. Both terms have components only in the out-of plane ($\hz$) direction and hence affect only the flexural modes. In-plane and out-of-plane modes are decoupled in the harmonic approximation in the sense that the dynamical matrix can be block diagonalized. However the modes {\it are} coupled through the elastic constant $\e_1$, since the dangling bond term $\cU_L^{\e_1}$ has both in-plane and out-of-plane components.  

\section{Results}
\begin{figure}
 \centering
 \includegraphics[width=6cm,bb=0 0 365 449]{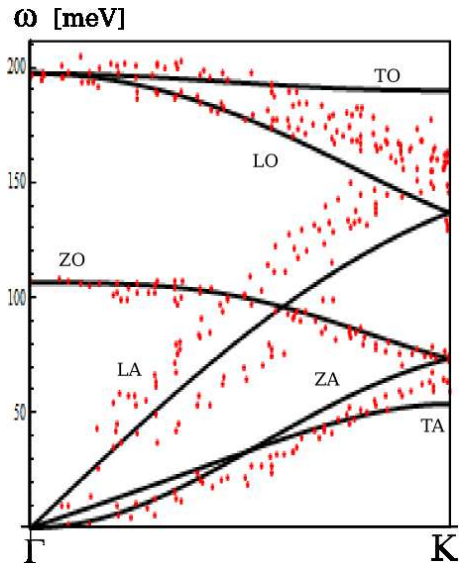}
 \caption{Solid line: phonon dispersion for the Lenosky model Vs. momentum with parameters as discussed in the text, dots: collection of experimental data taken from the review article Ref. \onlinecite{Rubio04}. The frequencies are in meV.}
 \label{fig:PhLenosky}
\end{figure}

In Lenosky's original work \cite{Lenosky92}, the value of $\e_0$ was taken to infinity and the remaining parameters were calculated by Local Density Approximation (LDA). The obtained values were $\e_1=0.96$ eV, $\e_2=1.29$ eV, $\e_3=0.05$ eV. Here we take the parameter $\e_0$ as adjustable to reproduce correctly the value of the optical modes at the $\Gamma$ point, from which we obtain $\e_0=36$ eV. In Fig. \ref{fig:PhLenosky} it is shown the phonon dispersion resulting from the Lenosky model plotted as a function of momentum with the above values for the $\e_n$ parameters.
As we anticipated, expression \eqref{UL} gives a very good description for the flexural modes near the $\Gamma$ point, as it can be seen from Fig. \ref{fig:PhLenosky}. However the description is not so good for the in plane modes. The failing of the model for the LO mode at the $\Gamma$ point and the transverse optic (TO) mode at the $K$ point can be attributed to the Kohn anomalies \cite{Piscanec04, Lanzara08}, that are not taken into account in \eqref{UL}. However the most important qualitative failure of the model is for the ordering of the modes at the $K$ point. With Lenosky's energy, the ZA/ZO modes at the $K$ point have larger energy than the transverse acoustic (TA) mode, on the contrary of what is observed experimentally. A quick inspection at the expressions of these modes in terms of the parameters $\e_n$ reveals that this problem cannot be solved by choosing a different set of parameters. The frequencies of the ZA/ZO and the TA modes at the $K$ point are given by $\w_{ZA}(K)=1/a\sqrt{\l(18 \e_1 +12 \e_2\rg)/M}$ and $\w_{TA}(K)=1/a\sqrt{18 \e_1/M}$ respectively, being $M$ the carbon mass, and where we have used the fact that $\e_0$ has to be the largest energy scale in the problem. Since $\e_n>0$, from these expressions it is evident that $\w_{ZA}(K)>\w_{TA}(K)$.

To overcome this issue we construct a ``hybrid'' model by adding to Lenosky's energy \eqref{UL} a bond-bending term as present in Keating and Kirkwood models:
\beq
\cU_B=\sum_{i=1}^{N}\sum_{\a,\;\nu>\a}\b \l(\br_{i\a(i)}\cdot \br_{i\nu(i)}+\half\rg)^2,
\eeq
which represents the energy cost of changing the angle between bonds. This is roughly equivalent then to taking the model presented in Ref. \onlinecite{Lobo97} and adding the $\pi$-$\pi$ and $\pi$-$\s$ overlap terms, besides the small difference in the dangling bond term discussed previously. The harmonic expansion $\cU_B^\b$ of this term can be readily obtained and is given by
\beq
\cU_B^\b= \sum_{X\neq Y}\sum\l[\iuno\cdot\l(\bu_{2(n)}^Y-\bu_n^X\rg) +\idos\cdot\l(\bu_{1(n)}^Y-\bu_n^X\rg)\rg]^2
\eeq
where the second sum is given over cyclic permutations of the nearest-neighbor indices. 

\begin{figure}
 \centering
 \includegraphics[width=7cm,bb=0 0 597 477]{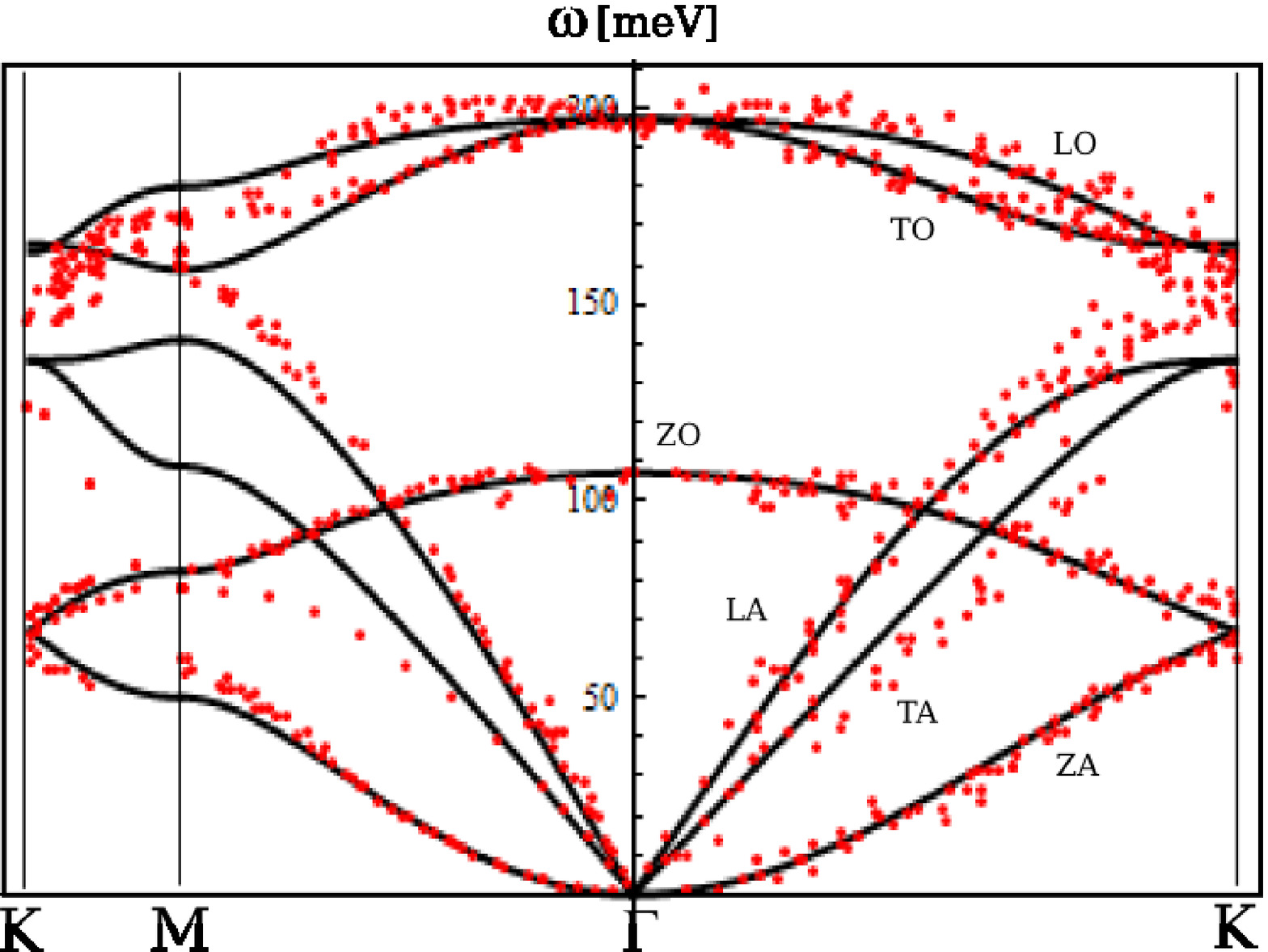}
 \caption{Solid line: best fit phonon dispersion for the hybrid model Vs. momentum with parameters as discussed in the text, dots: collection of experimental data taken from the review article Ref. \onlinecite{Rubio04} and Ref. \onlinecite{Maultzsch07}. The frequencies are in meV.}
 \label{fig:HybUnrFit}
\end{figure}

\begin{figure}
 \centering
 \includegraphics[width=7cm,bb=0 0 597 477]{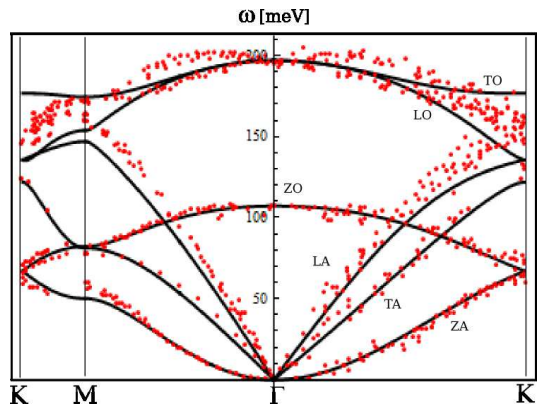}
 \caption{Solid line: phonon dispersion for the hybrid model Vs. momentum resulting for a restricted fit as discussed in the text, dots: collection of experimental data taken from the review article Ref. \onlinecite{Rubio04} and Ref. \onlinecite{Maultzsch07}. The frequencies are in meV.}
 \label{fig:HybResFit}
\end{figure}

We present the results from our hybrid model in Fig. \ref{fig:HybUnrFit} for the high symmetry lines of the full Brioullin zone. We obtained a best fit for the experimental data in the $\Gamma$-$K$ direction and utilized the resulting values for the elastic parameters $\e_0=24.8$ eV, $\e_1=1.3$ eV, $\e_2=0.2$ eV, $\e_3=1.2$ eV and $\b=5.0$ eV to construct the dispersion for the rest of the Brioullin zone.
From Fig. \ref{fig:HybUnrFit} it is seen that the fitting for the flexural modes is indeed exceptionally good. However problems still remain for the in-plane modes. In particular the degeneracy of the longitudinal acoustic (LA)/LO and TA/TO modes at the $K$ point is not the right one. This problem is a consequence of the bond-bending term $\cU_\b$ and it is also present in both nearest-neighbor Keating and Kirkwood models, although it is not mentioned in the literature. The way out of this problem is to realize that these models allow for the right symmetry if a condition among the elastic parameters is fullfilled. We found this condition for our hybrid model to be $\e_0 >\l(7 \b + \frac{9}{2}\e_1\rg)$ --- which reduces to $\e_0 >7 \b$ for the usual Keating model. Therefore, the fitting of the elastic parameters has to be constrained by this condition, instead of an unrestricted fit. If we apply this restriction to our fitting of the data, we obtain the phonon spectrum depicted in Fig. \ref{fig:HybResFit}.   
From the figure it can be seen that the new fit has the right symmetries and it is overall a very good fit in the whole Brioullin zone, with the exception of the points where the Kohn anomaly should play a role: namely the overbending of the LO mode at the $\Gamma$ point and the softening of the TO mode at the $K$ point. From the values obtained for the elastic constants $\e_0=30$ eV, $\e_1=1.3$ eV, $\e_2=0.2$ eV, $\e_3=1.2$ eV and $\b=2.4$ it can be inferred that the $\pi$-$\s$ overlap dominates over the $\pi$-$\pi$ orbital overlap. This is in agreement with the results presented in Ref. \onlinecite{Haas92} within a tight binding study but differ with the values obtained for the Lenosky energy in Ref. \onlinecite{Lenosky92} in which $\e_3$ is the smallest parameter.

\section{Conclusions}
In conclusion, we have analyzed the phonon dispersion given by the energy function \eqref{UL} first introduced in Ref. \onlinecite{Lenosky92} for graphene sheets. We have shown that this model gives good results for the flexural modes but fails to describe correctly in-plane modes beyond the $\Gamma$ point. To overcome this we have constructed a five parameters hybrid model which adds to Lenosky's energy a commonly used bond-bending term. We have shown that a restricted fit of the elastic parameters reproduces correctly all the relevant features of the phonon dispersion of graphene, with the exception of the Kohn anomalies. This restricted fit is essential to obtain the right symmetries of the model, and our results with respect to this point also apply for existent and well established nearest-neighbor valence force models in the literature, namely, the Kirkwood \cite{Kirkwood39} and Keating \cite{Keating66} models. According to our results, the effect of the change in overlap between the $\pi$ and $\s$ bonds is dominant over the effect due to the change in  the $\pi$-$\pi$ overlap. The agreement of our model with the experimental data is extremely good. In particular the flexural modes, which are of extreme importance for graphene, are excellently described. We have therefore obtained a minimal, few parameters model which can be used as a starting point in future analytical calculations. Results in this respect will be published elsewhere.

\section{Acknowledgments}
SVK thanks D. Guerra for helpful suggestions for the data analysis, and L. Malard, V. Pereira and A. Swan for fruitful discussions. AHCN acknowledges the partial support of the U.S. Department of Energy under grant DE-FG02-08ER46512.


\begin{thebibliography}{20}
\bibitem{note0} The dangling bond term introduced in Ref. \onlinecite{Lobo97} can be obtained from the Lenosky's one by doing the approximation $\br_{ij}/\l|\br_{ij}\rg|\approx\br_{ij}/r_0$. The former presents only two branches, optical and acoustical, which are three-fold degenerate. This degeneracy is completely lifted in Lenosky's approximation.
\bibitem{note1} Note that the honeycomb lattice consists of two inequivalent triangular sublattices $A$ and $B$ as depicted in Fig. \ref{fig:labels}.
\end{thebibliography}
\end{document}